\documentstyle [12pt] {article}

\catcode`@=12
\input epsf.tex
\def\DESepsf(#1 width #2){\epsfxsize=#2 \epsfbox{#1}}
\begin{document}
\thispagestyle{empty}
\begin{flushright} UCRHEP-T188\\July 1997\
\end{flushright}
\vspace{0.5in}
\begin{center}
{\Large	\bf S, T, and Leptoquarks at HERA\\}
\vspace{1.5in}
{\bf E. Keith and Ernest Ma\\}
\vspace{0.3in}
{\sl Department of Physics\\}
{\sl University of California\\}
{\sl Riverside, California 92521\\}
\vspace{1.5in}
\end{center}
\begin{abstract}\
If the recently discovered anomalous events at HERA are due to a scalar
leptoquark, then it is very likely to have weak isospin $I = 1/2$.  In that
case, present precision measurements of the oblique radiative parameters
$S$ and $T$ provide strong constraints on the mass of the other component of
this doublet.  If the standard model is extended to include such a doublet,
a slightly better fit may in fact be obtained.  However, in specific
proposed models where this doublet comes from a larger symmetry, there
are often additional large and positive contributions to $S$ from exotic
heavy fermions which far exceed the present experimental limit.  A way
to improve the Tevatron exploration of leptoquarks is proposed.
\end{abstract}

\newpage
\baselineskip 24pt

Recently, two experiments at the HERA accelerator have observed some
anomalous events in $e^+ p$ scattering,\cite{1,2} which may be
interpreted as due to a scalar leptoquark of mass about 200 GeV.\cite{3}
Assuming that a constituent quark of the proton is involved, then there are
only four possibilities for this scalar leptoquark (call it $\eta$)
according to what it is coupled to:
\begin{equation}
(1) ~e^+_L u_L, ~~~(2) ~e^+_L d_L, ~~~(3) ~e^+_R u_R, ~~~(4) ~e^+_R d_R.
\end{equation}
We find it more convenient to rewrite the last two cases in terms
of their Hermitian conjugates:
\begin{equation}
(3) ~e^-_L \bar u_L, ~~~(4) ~e^-_L \bar d_L.
\end{equation}
It is thus obvious that $\eta$ has weak isospin $I=1/2$ with $I_3 = 1/2, -1/2,
-1/2, -1/2$ and weak hypercharge $Y = 7/6, 7/6, -7/6, -1/6$ respectively.
The first two combine to form a doublet, whereas the last two have $I_3 = 1/2$
partners $\nu_L \bar u_L$ and $\nu_L \bar d_L$ respectively.

Let us assume that the standard $SU(3) \times SU(2) \times U(1)$ model of
quarks and leptons is extended to include one such scalar leptoquark doublet.
Its contribution to the oblique radiative parameters\cite{4} $S$ and $T$ are
easily calculated.\cite{5}  First, we have
\begin{equation}
\Delta S = {{-Y} \over {2 \pi}} \log {m_1^2 \over m_2^2},
\end{equation}
where $m_{1,2}$ are the masses of the $I_3 = \pm 1/2$ components of $\eta$.
Note that $\Delta S$ can be positive or negative, depending on $Y$ and
$m_1^2/m_2^2$.  This is in contrast to the well-known case of a fermion
doublet in the standard model, because there the left-handed components
form an $SU(2)$ doublet but the righthanded components are singlets,
resulting in a positive value of $1/6 \pi$ for each such doublet when
$m_1 = m_2$.  This fact is a very important constraint on models beyond
the standard model and we will come back to it later.

Second, we have
\begin{equation}
\Delta T = {3 \over {16 \pi}} {1 \over {s^2 c^2 M_Z^2}} \left[ m_1^2 + m_2^2
- {{2 m_1^2 m_2^2} \over {m_1^2 - m_2^2}} \log {m_1^2 \over m_2^2} \right],
\end{equation}
where $s^2 \equiv \sin^2 \theta_W$ and $c^2 \equiv \cos^2 \theta_W$.  As
expected, $\Delta T$ is necessarily nonnegative, and is zero only if
$m_1 = m_2$.  A recent analysis\cite{6} of all relevant experimental data
obtained the following values for $S$ and $T$:
\begin{equation}
S = -0.11 \pm 0.13 \begin{array} {c} -0.09 \\ +0.08 \end{array}, ~~~
T = -0.03 \pm 0.14 \begin{array} {c} +0.17 \\ -0.12 \end{array},
\end{equation}
where $m_t = 175$ GeV, $\alpha_S = 0.118$, and the second set of
uncertainties corresponds to the choice of the
Higgs-boson mass: 1 TeV for the upper value, 300 GeV for the central value,
and 100 GeV for the lower value.  The standard-model contributions have
been subtracted, so the above numbers represent how much new physics
contributions are to be tolerated.  Another recent analysis\cite{7}
considers the parameters $\epsilon_{1,2,3}$\cite{8} with results consistent
with the above.

The addition of a scalar doublet of leptoquarks to the standard model will
in general shift $S$ and $T$.  If the two components are degenerate in mass,
the shifts do vanish in lowest order.\cite{9}  This may well be the case
with the HERA data.  It means that for the $Y = 7/6$ interpretation,
both components of the scalar leptoquark are produced, whereas for the
$Y = -7/6$ or $-1/6$ interpretation, there are also $I_3 = 1/2$ partners with
the same mass ({\it i.e.} 200 GeV) which decay into $\nu_L \bar u_L$ and
$\nu_L \bar d_L$ respectively.  Since all leptoquarks are also accessible
at hadron colliders, these latter decay modes should also be searched for.

If the scalar leptoquark is not degenerate in mass, then $\Delta T$ will be
positive, but $\Delta S$ can be either positive or negative.  It is clear
from the data that $\Delta S < 0$ is preferred.  Now if we make the
reasonable assumption that whereas the HERA leptoquark is 200 GeV in mass,
its doublet partner ought to be heavier, then to obtain a negative $\Delta S$,
Case (2) should be chosen instead of Case (1).  As for Cases (3) and (4),
$\Delta S < 0$ implies that the missing partner should be lighter and
decays into $\nu_L \bar u_L$ and $\nu_L \bar d_L$ as already mentioned.  For
Cases (2) and (3), it is in fact possible to have a slightly better fit to
the data than with the standard model.

In Figure 1, we plot $\Delta T$ of Eq.~(4) as a function of $m_1$, assuming
$m_2 = 200$ GeV.  Since this expression is symmetric with respect to the
interchange of $m_1$ and $m_2$, it also applies to fixing $m_1$ at 200 GeV
and varying $m_2$.  The experimental upper limit on $T$ tells us that
$|m_1 - m_2|$ cannot be too large.  In Figure 2, we plot $\Delta S$ of
Eq.~(3) as a function of $m_1$, assuming $m_2 = 200$ GeV and $Y = 7/6$,
{\it i.e.} Case (2).  The experimental preference for a negative $S$ tells
us that $m_1 > m_2$ is more likely in this case.  Finally in Figures 3(a) and
3(b), we show the locus of points corresponding to Case (2) in the $S-T$ plane
for $m_H = 100$ and 300 GeV respectively.  In 3(a), the point on this curve
closest to the center of the experimentally determined ellipse\cite{6} 
(defined by
the sum of its distances to the two foci) corresponds to $m_1 \sim 200$ GeV,
{\it i.e.} doublet degeneracy.  The curve intersects the 90(99)\%
confidence-level (CL) ellipse at $m_1 \sim 140(120)$ GeV and 240(260) GeV.
In 3(b), closest approach corresponds to $m_1 \sim 220$ GeV, and the limits
on $m_1$ are 130(110) GeV and 270(280) GeV at the 90(99)\% CL.

It is natural to expect $m_1 \neq m_2$ for the scalar leptoquark doublet
$\eta \equiv (\eta_1, \eta_2)$ as shown below.  The scalar potential here
consists of the usual $\Phi \equiv (\phi^+, \phi^0)$ doublet of the
standard model as well as $\eta$, {\it i.e.}
\begin{equation}
V = \mu^2 \Phi^\dagger \Phi + m^2 \eta^\dagger \eta + {1 \over 2} \lambda_1
(\Phi^\dagger \Phi)^2 + {1 \over 2} \lambda_2 (\eta^\dagger
\eta)^2 + \lambda_3 (\eta^\dagger \eta)(\Phi^\dagger \Phi) + \lambda_4
(\eta^\dagger \Phi)(\Phi^\dagger \eta).
\end{equation}
With $\langle \phi^0 \rangle = v$, it is easily obtained from the above that
\begin{equation}
m_1^2 = m^2 + \lambda_3 v^2, ~~~ m_2^2 = m^2 + (\lambda_3 + \lambda_4) v^2.
\end{equation}
Hence $m_1^2 - m_2^2 = -\lambda_4 v^2$ and is in general nonzero and may be
of either sign.

We are aware of two models which predicted scalar leptoquarks corresponding
to one or more of the four cases being studied.  One\cite{10} is based on
$SU(15)$ and essentially predicts all possible leptoquark combinations.
If correct, many other scalar leptoquarks will be found, both in $e^+ p$
and $e^- p$ scattering.  It also predicts dileptons and diquarks.  The
other\cite{11} is based on $SO(10) \times U(1)$ and predicts only one scalar
leptoquark doublet corresponding to $Y = 7/6$.  It also predicts a number
of exotic particles, but none in the $e^- p$ channel.  However, both
models suffer from a large and positive contribution to $S$.  In the former,
because of the need for anomaly cancellation, there are three families of
mirror fermions, hence $S$ receives a contribution of $2/\pi (= 0.64)$.  In
the latter, because the model is supersymmetric and there are two additional
exotic families, the contribution is $4/3\pi (= 0.42)$.  In the above, we have
assumed doublet degeneracy so that $\Delta T = 0$.  Relaxing that assumption
only makes the disagreement with data worse.

A supersymmetric model based on $SU(3)_C \times SU(2)_L \times SU(2)_R
\times U(1)_{B-L}$ proposed recently\cite{12} assigns the HERA scalar
leptoquarks to the representations $(3, 2, 2, 4/3)$ and $(3^*, 2, 2, -4/3)$.
Their fermion partners have gauge-invariant masses, hence they do not
contribute to $S$.  However, if these supermultiplets are embedded into
a larger symmetry such as $SU(5) \times SU(5)$ proposed earlier\cite{13},
there will be in general positive contributions to $S$.  For example, in the
supermultiplets $({\bf 5}, {\bf 10}) + (\bar {\bf 10}, \bar {\bf 5})$, there
are unpaired fermion multiplets $(6^*, 2, 1, 1/3)$ and $(1, 2, 1, 3)$
which would contribute $7/6\pi (= 0.37)$ to $S$.

The constraint of $S$ on unpaired fermion doublets is important in other
models as well.  For example, in a specific $SO(10) \times SO(10)$
model,\cite{14} there is an extra $(\bar {\bf 16}, {\bf 1})$ supermultiplet,
hence $S$ receives a contribution of $2/3\pi (= 0.21)$.  On the other hand,
in supersymmetric $E_6$ models, the {\bf 27} representation is safe in
this respect because it contains no additional unpaired doublet beyond those
of the standard model.

If the HERA anomalous events are due to scalar leptoquarks, then they
can be produced in proton-antiproton collisions at Fermilab.  However,
preliminary reports\cite{15} from the CDF and D0 experiments at the Tevatron
indicate that they are excluded up to 210 and 225 GeV in mass respectively.
Since such a leptoquark should be a member of a doublet and
according to the constraints from $S$ and $T$, if one mass is 200 GeV,
the other should be close to it, both should be produced at the Tevatron.
For example if $m_1 = m_2$, then both $\eta_1$ and $\eta_2$ should be
produced, thereby doubling the putative cross section, resulting in an
even more stringent bound on the leptoquark mass.

In conclusion, if a scalar leptoquark is responsible for the anomalous
$e^+ p$ events at HERA, then it is very likely to have weak isospin $I = 1/2$.
The constraints from precision measurements at the $Z$ pole tells us that
if one component of this doublet is at 200 GeV, the other cannot be too
far away.  This allows for a very stringent test of leptoquarks at the
Tevatron.  At present, assuming the production of a single scalar leptoquark,
preliminary lower limits of 210 and 225 GeV are being reported by the CDF and
D0 collaborations.  If one assumes the production of two scalar leptoquarks,
then a lower limit can be placed upon their masses in a two-dimensional plot.
This can be done by assuming a fixed ratio of $m_1/m_2$ with $m_1 > m_2$, 
then determining the lower limit on $m_2$.  If $m_1/m_2$ is large, the 
present bounds for the production of a single scalar leptoquark are recovered. 
In Cases (1) and (2), the decays of both components of the doublet are into
$e^+ q$ (or $e^- \bar q$).  In Cases (3) and (4), one component decays into
$\bar \nu q$ (or $\nu \bar q$) which would involve large misssing energy.
Such an analysis will improve the present Tevatron bounds and may help to
rule out (or confirm) the leptoquark hypothesis.

\vspace{0.3in}
\begin{center} {ACKNOWLEDGEMENT}
\end{center}

We thank Paul Frampton, Seiji Matsumoto, and Peter Renton 
for correspondence.
This work was supported in part by the U. S. Department of Energy under
Grant No. DE-FG03-94ER40837.

\newpage
\bibliographystyle{unsrt}

\begin{thebibliography}{99}
\bibitem{1} H1 Collaboration, C. Adloff {\it et al.}, Z. Phys. {\bf C74}, 
191 (1997).
\bibitem{2} ZEUS Collaboration, J. Breitweg {\it et al.}, Z. Phys. {\bf C74}, 
207 (1977).
\bibitem{3} Many theoretical papers have appeared since the HERA
experimental announcements.  For a short summary, see for example P. H.
Frampton, hep-ph/9706220.
\bibitem{4} M. E. Peskin and T. Takeuchi, Phys. Rev. Lett. {\bf 65}, 964
(1990).
\bibitem{5} See for example C. D. Froggatt, R. G. Moorhouse, and I. G.
Knowles, Phys. Rev. {\bf D45}, 2471 (1992).
\bibitem{6} K. Hagiwara, D. Haidt, and S. Matsumoto, hep-ph/9706331.
\bibitem{7} P. Renton, Oxford Univ. Report OUNP-97-01, Int. J. Mod. Phys.
{\bf A}, to be published.
\bibitem{8} G. Altarelli, R. Barbieri, and S. Jadach, Nucl. Phys. {\bf B369},
3 (1992); {\bf B376}, 444(E) (1992).
\bibitem{9} J. L. Hewett and T. G. Rizzo, hep-ph/970337.  See also J. K.
Mizukoshi, O. J. P. Eboli, and M. C. Gonzalez-Garcia, Nucl. Phys.
{\bf B443}, 20 (1995).
\bibitem{10} P. H. Frampton, Mod. Phys. Lett. {\bf A7}, 559 (1992).
\bibitem{11} V. Barger and E. Ma, Phys. Rev. {\bf D51}, 1332 (1995).
\bibitem{12} B. Dutta, R. N. Mohapatra, and S. Nandi, hep-ph/9704428.
\bibitem{13} R. N. Mohapatra, Phys. Lett. {\bf B379}, 115 (1996); Phys. Rev.
{\bf D54}, 5728 (1996).
\bibitem{14} E. Ma, Phys. Rev. {\bf D51}, 236 (1995).
\bibitem{15} See for example {\em Fermi News} (June 20, 1997).

\end{thebibliography}

\newpage
\begin{center} {\large \bf Figure Captions}
\end{center}

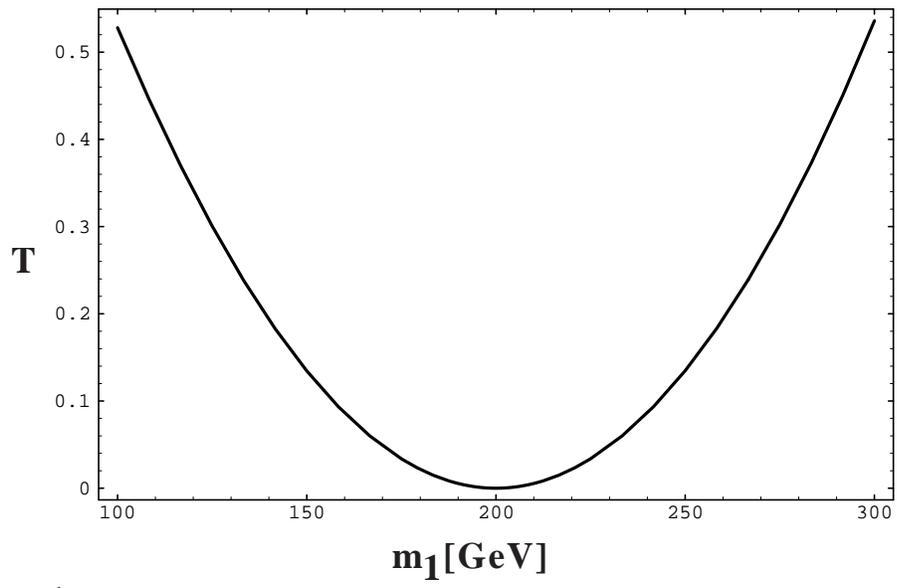
\noindent Fig.~1.  Contribution to $T$ from the scalar leptoquark doublet 
given by Eq.~(4) as a function of $m_1$ for $m_2 = 200$ GeV.

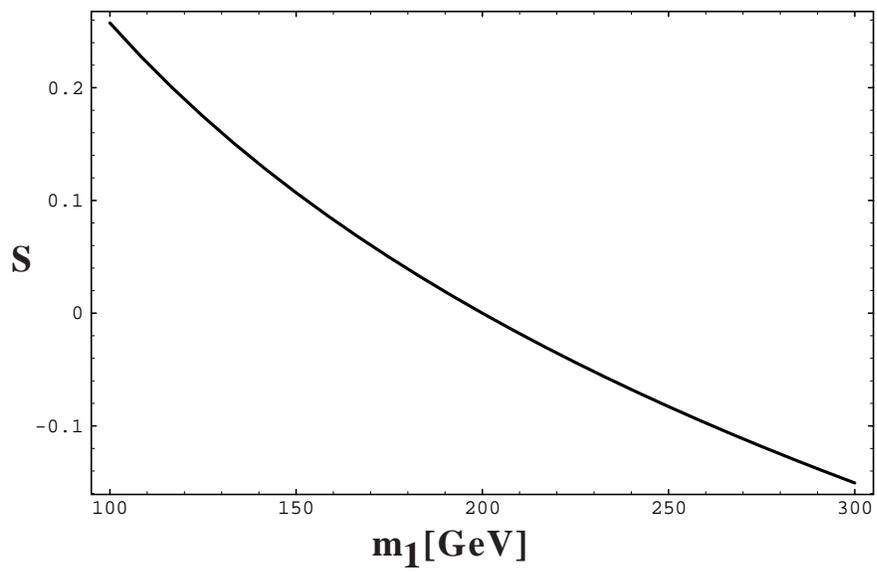
\noindent Fig.~2.  Contribution to $S$ from the scalar leptoquark doublet 
given by Eq.~(3) as a function of $m_1$ for $m_2 = 200$ GeV and $Y = 7/6$, 
{\it i.e.} Case (2).
 
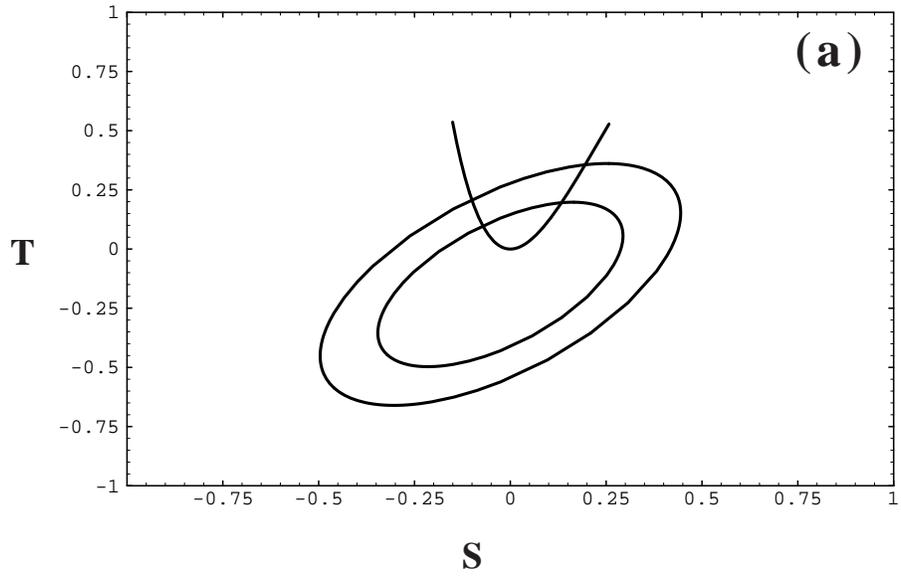
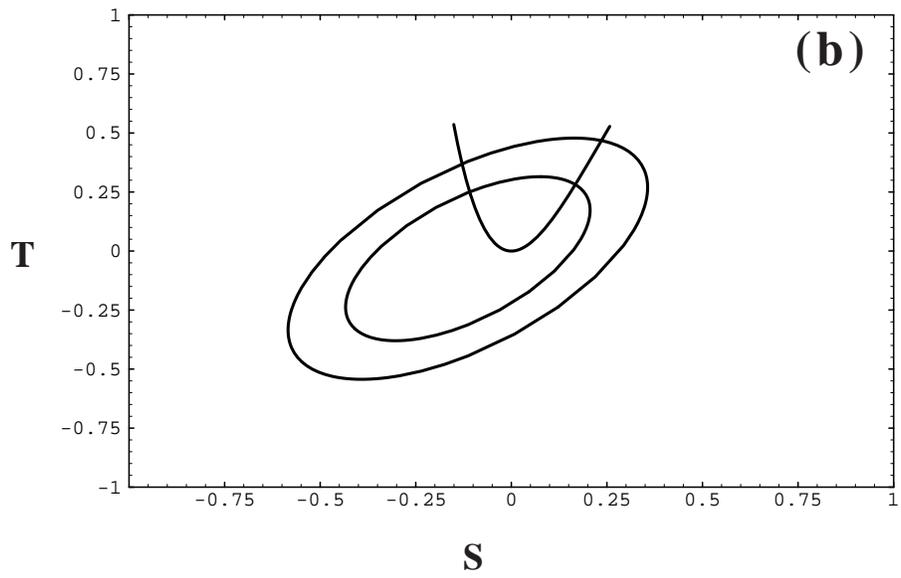
\noindent Fig.~3.  (a):  Parameteric plot of the Case (2) ($m_2=200$ GeV 
and $Y=7/6$) $T$ and $S$ values of Fig.~1 and Fig.~2 for the range 
100 GeV $ < m_1 < $ 300 GeV together with the experimentally determined 90\%
(inner) and the 99\% CL (outer) ellipses for $T$ and $S$ from new physics 
when it is assumed that $m_t = 175$ GeV and $m_H = 100$ GeV according to 
Ref.~[6]. (b): Same as (a) but the ellipses correspond to $m_H=300$ GeV.

\newpage
\begin{figure}[htb]
\centerline{ \DESepsf(f1f2.epsf width 15 cm) } \smallskip
\nonumber
\end{figure}

\newpage
\begin{figure}[htb]
\centerline{ \DESepsf(f3.epsf width 15 cm) } \smallskip
\nonumber
\end{figure}

\end{document}